\documentclass[aps,twocolumn,superscriptaddress]{revtex4-1}
\usepackage{amsfonts}
\usepackage{dcolumn}
\usepackage{bm}
\usepackage{tikz}
\usepackage[colorlinks=true,citecolor=blue,urlcolor=blue]{hyperref}
\usepackage{amsmath,amsfonts,amssymb,times,natbib}
\usepackage[standard]{ntheorem}

\begin{document}

\title{Topological band structure via twisted photons in a degenerate cavity}

%%%%%%%%%%%%%%%%%%%%%%%%%%%%%%%%%%%%%%%%%%%%%%%%%%%%%%%%%%%%%%%%%%%

\author{Mu Yang \footnote{These authors contribute equally to this work\label{Contribute}}}
\thanks{These authors contribute equally to this work\label{Contribute}}
\affiliation{CAS Key Laboratory of Quantum Information, University of Science and Technology of China, Hefei 230026, People's Republic of China}
\affiliation{CAS Center For Excellence in Quantum Information and Quantum Physics, University of Science and Technology of China, Hefei 230026, People's Republic of China}

\author{Hao-Qing Zhang\textsuperscript{\ref{Contribute}}}
\affiliation{CAS Key Laboratory of Quantum Information, University of Science and Technology of China, Hefei 230026, People's Republic of China}
\affiliation{CAS Center For Excellence in Quantum Information and Quantum Physics, University of Science and Technology of China, Hefei 230026, People's Republic of China}

\author{Yu-Wei Liao\textsuperscript{\ref{Contribute}}}
\affiliation{CAS Key Laboratory of Quantum Information, University of Science and Technology of China, Hefei 230026, People's Republic of China}
\affiliation{CAS Center For Excellence in Quantum Information and Quantum Physics, University of Science and Technology of China, Hefei 230026, People's Republic of China}

\author{Zheng-Hao Liu}
\affiliation{CAS Key Laboratory of Quantum Information, University of Science and Technology of China, Hefei 230026, People's Republic of China}
\affiliation{CAS Center For Excellence in Quantum Information and Quantum Physics, University of Science and Technology of China, Hefei 230026, People's Republic of China}

\author{Zheng-Wei Zhou}
\affiliation{CAS Key Laboratory of Quantum Information, University of Science and Technology of China, Hefei 230026, People's Republic of China}
\affiliation{CAS Center For Excellence in Quantum Information and Quantum Physics, University of Science and Technology of China, Hefei 230026, People's Republic of China}

\author{Xing-Xiang Zhou}
\affiliation{CAS Key Laboratory of Quantum Information, University of Science and Technology of China, Hefei 230026, People's Republic of China}
\affiliation{CAS Center For Excellence in Quantum Information and Quantum Physics, University of Science and Technology of China, Hefei 230026, People's Republic of China}

\author{Jin-Shi Xu}\email{jsxu@ustc.edu.cn}
\affiliation{CAS Key Laboratory of Quantum Information, University of Science and Technology of China, Hefei 230026, People's Republic of China}
\affiliation{CAS Center For Excellence in Quantum Information and Quantum Physics, University of Science and Technology of China, Hefei 230026, People's Republic of China}

\author{Yong-Jian Han}\email{smhan@ustc.edu.cn}
\affiliation{CAS Key Laboratory of Quantum Information, University of Science and Technology of China, Hefei 230026, People's Republic of China}
\affiliation{CAS Center For Excellence in Quantum Information and Quantum Physics, University of Science and Technology of China, Hefei 230026, People's Republic of China}
	
\author{Chuan-Feng Li}
\email{cfli@ustc.edu.cn}
\affiliation{CAS Key Laboratory of Quantum Information, University of Science and Technology of China, Hefei 230026, People's Republic of China}
\affiliation{CAS Center For Excellence in Quantum Information and Quantum Physics, University of Science and Technology of China, Hefei 230026, People's Republic of China}

\author{Guang-Can Guo}
\affiliation{CAS Key Laboratory of Quantum Information, University of Science and Technology of China, Hefei 230026, People's Republic of China}
\affiliation{CAS Center For Excellence in Quantum Information and Quantum Physics, University of Science and Technology of China, Hefei 230026, People's Republic of China}

\begin{abstract}
    {Synthetic dimensions based on particles' internal degrees of freedom, such as frequency, spatial modes and arrival time, have attracted significant attention. They offer ideal large-scale lattices to simulate nontrivial topological phenomena. Exploring more synthetic dimensions is one of the paths toward higher dimensional physics.
    In this work, we design and experimentally control the coupling among synthetic dimensions consisting of the intrinsic photonic orbital angular momentum and spin angular momentum degrees of freedom in a degenerate optical resonant cavity, which generates a periodically driven spin-orbital coupling system. We directly characterize the system's properties, including the density of states, energy band structures and topological windings, through the transmission intensity measurements. 
    Our work demonstrates a novel mechanism for exploring the spatial modes of twisted photons as the synthetic dimension, which paves the way to design rich topological physics in a highly compact platform.}
\end{abstract}

\date{\today}

\maketitle

%%%%%%%%%%%%%%%%%%%%%%%%%%%%%%%%%%%%%%%%%%%%%%%%%%%%%%%%%%%%%%%%%%%%%%%%%%%%%%%%%%%%%%%%%%%%%%%%%%%%%%%%%%%%%%%%%%%%%%%%%%%%%%%%%
%\noindent{\bf \large Introduction}\\
%%%%%%%%%%%%%%%%%%%%%%%%%%%%%%%%%%%%%%%%%%%%%%%%%%%%%%%%%%%%%%%%%%%%%%%%%%%%%%%%%%%%%%%%%%%%%%%%%%%%%%%%%%%%%%%%%%%%%%%%%%%%%%%%%
%\textit{Introduction}.---
%Spin-orbit coupling (SOC) usually play an important role in non-trivial topological phenomenons.
\section{Introduction}
The dimensions of physical models simulated by real space lattices, such as photonic crystals~\cite{photoncry,photoncry1,photoncry2}, metamaterials~\cite{mate} and microcavity arrarries~\cite{cavityarray}, are generally smaller than or equal to their geometric dimensions. Great efforts have been made to simulate high-dimensional physics.
%It is still a big challenge to experimentally simulate the physics in high dimensions. 
Recently, a powerful approach by introducing synthetic dimensions to the lower geometric dimensions with remarkably fewer experimental requirements has caused increasing interest~\cite{Fan1, high3}. The $(D+d)$-dimensional physics can be investigated in $D$ geometric dimensions with $d$ synthetic dimensions.
The synthetic dimensions could be formed by the particles' internal degrees of freedom in photonic~\cite{Fan1,high2} or atomic systems~\cite{spin,spin2,spin3}.

Abundant topological phenomena have been demonstrated through utilizing the photonic frequencies~\cite{Fane2,Fane3,Fane1,Fane4}, optical waveguide modes~\cite{synthetic1} and optical pulse arrival time~\cite{plus1,plus0} as synthetic physical dimensions. Exploring more synthetic dimensions is helpful to investigate higher dimensional physics.

Photonic orbital angular momentum (OAM) with infinite topological charge numbers is an ideal degree of freedom for constructing the synthetic lattice. The photons carrying OAM have twisted wavefronts, referred to as twisted photons~\cite{twist}. Moreover, optical systems with a tunable coupling between the intrinsic spin angular momentum (SAM) and the synthetic OAM dimension of photons offer natural platforms to simulate the topological physics in spin-orbital coupling (SOC) systems. As the first proposal for synthetic dimensions~\cite{high2,high3}, degenerate optical cavities simultaneously support plenty of OAM modes have been employed in theoretical protocols to simulate a wide variety of topological physics including the non-Abelian gauge fields induced phase transition ~\cite{Luo1} and edge states~\cite{Luo3}. The synthetic frequency and OAM dimensions are combined in a single cavity to investigate gauge field physics~\cite{Fan3}. Multimode optical cavities have also been experimentally used to simulate Landau levels~\cite{caviye2, caviye3}, which are the first two-dimensional topological cavity, and have also been used to engineer Hamiltonians~\cite{ec}. %\red{However, it is still abcence in experimental demonstration of synthetic OAM dimensions in a cavity.} 
However, there has been no experimental demonstration of the OAM degenerate cavity-assisted SOC physics until now.

In this work, we develop an extremely compact platform and experimentally investigate the properties of a periodically driven SOC topological system in a degenerate optical cavity with the photonic OAM serving as a synthetic dimension. The coupling between the synthetic dimension and its internal spin is well constructed. By detecting the transmission intensity of the degenerate cavity, we directly obtain the density of states (DOS), energy band structures, and topological windings of the simulated system. Although some topological evidences have been demonstrated through quantum walk on OAM modes in stackable systems~\cite{QWe1,QWe3,QWe2}, the direct band observations are not available. Our experiments open the door to directly explore high-dimensional topological physics with synthetic dimensions in a simple system.

\section{Results}
\subsection{Theoretical framework}
\begin{figure*}[tbp]
	\centering
	\includegraphics[width=1.8\columnwidth]{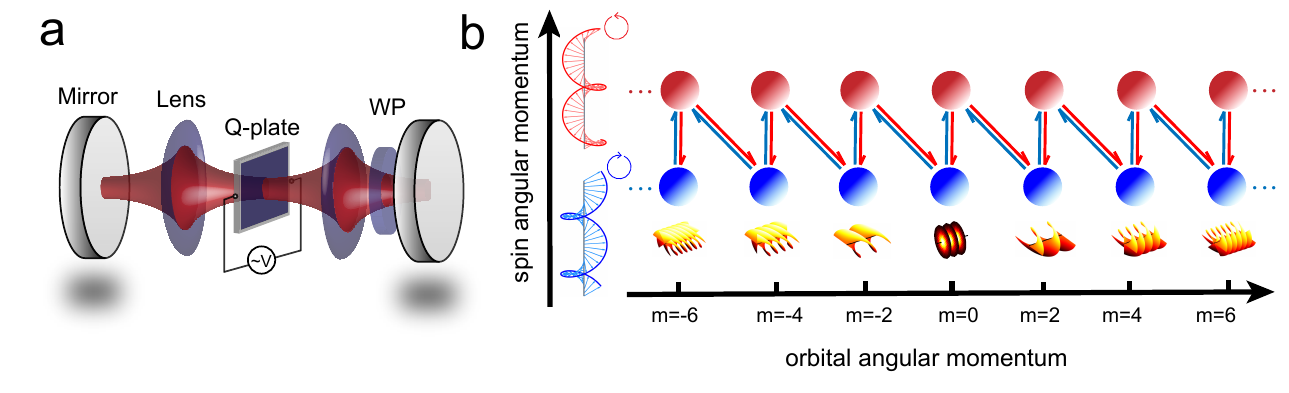}
	\caption{\label{concept}\small{
	\textbf{Experimental setup and the coupling model}. (\textbf{a}). The degenerate cavity consists of two plane mirrors and two lenses. High-order photonic modes are generated and coupled by repeatedly passing an anisotropic and inhomogeneous medium (Q-plate) and a wave plate (WP) with an input Gaussian mode. (\textbf{b}). Schematic of spin and lattice for SAM and OAM modes. The SAM modes with left ($\circlearrowleft$) and right ($\circlearrowright$) circular polarizations are labeled in red and blue, respectively. The OAM modes are marked as the arrays of balls. The corresponding twisted wavefronts are shown below the coupling modes. 
	}}
\end{figure*}

The degenerate optical cavity shown in FIG. \ref{concept}a consists of two high reflective plane mirrors and two convex lenses. The input optical mode locates one focal length (f) in front of the first lens while the output optical mode locates one focal length after the second lens, which forms an exact 4f system. The cavity can stabilize more than $10^3$ photonic OAM modes (see section I of Supplementary Information (SI) for more details), which steadily reproduce themselves periodically since they go through the precise 4f telescope once every period. 
A large synthetic lattice based on OAM modes can then be constructed in the optical degenerate cavity. The degenerate cavity carrying a variety of OAM modes requires the precise control of surface quality and position of cavity elements. Aberrations may destroy its degenerate~\cite{caviye4, caviye5,caviye6}, which implies that the experimental requirement is high. We pursue this kind of degeneracy to reduce the self-energy disorder in the cavity which may destroy the topological character in the system.

The generation of high-order spatial OAM modes with the input Gaussian mode and the coupling between the OAM and SAM modes are simultaneously achieved via an anisotropic and inhomogeneous medium (named Q-plate~\cite{Qp}) in the cavity (see Methods). The action of the Q-plate ($J_{Q(\delta)}$) is described as:
 \begin{equation}
 \begin{aligned}
 J_{Q(\delta)}=\sum_{m}\cos(\delta/2)(a_{\circlearrowleft,m}^{\dag}a_{\circlearrowleft,m}+a_{\circlearrowright,m}^{\dag}a_{\circlearrowright,m})\\
 +i\sin(\delta/2)(a_{\circlearrowright,m+2q}^{\dag}a_{\circlearrowleft,m}+\mathrm{h.c.}),
 \end{aligned}
 \end{equation}
where $m$ represents the topological charge numbers of OAM modes with corresponding twisted wavefronts; $\circlearrowleft (\circlearrowright)$ denotes the left (right)-circular polarized SAM modes; $a^{\dag}_{\circlearrowleft(\circlearrowright),m}$ ($a_{\circlearrowleft(\circlearrowright),m}$) is the corresponding creation (annihilation) operator; $q$ is the topological charge number of the Q-plate and $q=1$ in our experiment; and $\delta$ is the value of the optical retardation which can be tuned by the applied electric field. To manipulate the optical SAM modes, an additional wave plate (WP) with the operation of
$
J_{\lambda(\eta)}=\sum_{m}\cos(\eta)(a_{\circlearrowleft,m}^{\dag}a_{\circlearrowleft,m}+a_{\circlearrowright,m}^{\dag}a_{\circlearrowright,m})
+i\sin(\eta)(a_{\circlearrowright,m}^{\dag}a_{\circlearrowleft,m}+\mathrm{h.c.})$ is introduced in the cavity. The control parameter $\eta$ represents the phase retardance between ordinary and extraordinary photons, which is determined by the WP's thickness (e.g. $\eta=\pi/4$ for a quarter-wave plate (QWP)). When the Q-plate and WP are introduced in the cavity,  the position and orientation of the cavity should be re-optimized to maintain the degenerate property.

The optical state in the cavity is evolved under the periodic unitary:
%$\hat{U}(\delta,\eta)=J_{Q(\delta)}J_{\lambda(\eta)}J_{\lambda(\eta)}J_{Q(\delta)}$.
$\hat{U}(\delta,\eta)=J_{Q(\delta)}J_{\lambda(\eta)}J_{\lambda(\eta)}J_{Q(\delta)}$ 
which is an one round trip including both the actions of the Q-plate and WP. Due to its periodicity, an effective Hamiltonian $\hat{H}_{\rm eff}$ can be introduced as $\hat{U}(\delta,\eta)=e^{-i\hat{H}_{\rm eff}T/\hbar}$, where $T=L/c$, with $L$ being the one round trip (one period) length of the cavity and $c$ representing the speed of light, denotes the period of a round-trip. The average lifetime of photons in the cavity is about $5T$, which means the photons pass through the Q-plate 10 times on average (see section I of SI for more details). 
The operation of $\hat{U}(\delta,\eta)$ drives the hopping among SAM and OAM modes (shown in FIG. \ref{concept}b), which shares the features of the simplest topological lattice Su-Schrieffer-Heeger (SSH) model~\cite{SSH} in the Floquet version. As a result, the stable optical state $\left|\phi(t)\right\rangle$ at time $t$, which is a superposition state of SAM and OAM modes ($\left|\circlearrowleft(\circlearrowright),m\right\rangle$), is evolved as $\left|\phi(t+T)\right\rangle=e^{-i\hat{H}_{\rm eff}T/\hbar}\left|\phi(t)\right\rangle$.
From the point of view of the self-reproductive condition, the stable optical states in the degenerate cavity satisfies: $\left|\phi(t+T)\right\rangle=e^{-i\beta L}\left|\phi(t)\right\rangle$. $\beta=2\pi/\lambda+i\alpha$ is independent of SAM and OAM modes. $\lambda$ represents the wavelength of the photons in the cavity and $\alpha$ is the attenuation coefficient.
Combining  the evolution and the reproductive conditions of the optical state in the degenerate cavity, we obtain
 \begin{equation}
e^{-i\hat{H}_{\rm eff}T/\hbar}\left|\phi(t)\right\rangle=e^{-i\beta L}\left|\phi(t)\right\rangle.
 \end{equation}
As a consequence, the stable photonic state $|\phi\rangle$ ($t$ is omitted) in the degeneracy cavity is naturally the eigenstates of $\hat{H}_{\rm eff}$ with eigenvalues $\beta L$ (taking $T/\hbar=1$).

Since the Q-plate and WP have same operational forms on different $m$, there should not be disorders in the coupling and the effective Hamiltonian $\hat{H}_{\rm eff}$ possesses translational symmetry on $m$. 
As a result, if we introduce the Bloch mode $|k\rangle$ in `momentum' space as $\left|k\right\rangle=\sum_{j}e^{-ijk}\left|j\right\rangle$ $(j=m/2)$, the Hamiltonian can be recast in the `quasi-momentum' space as $\hat{H}_{\rm eff}=\int^{\pi}_{-\pi}\hat{H}_{\rm eff}(k)dk$, where $\hat{H}_{\rm eff}(k)=E_{k}\textbf{n}(k)\cdot\bm{\sigma}\left|k\right\rangle\left\langle k\right|$. $E_{k}$ represents the dispersion relation, $\bm{\sigma}=(\sigma_{x},\sigma_{y},\sigma_{z})$ is the Pauli vector and $\textbf{n}(k)=[n_{x}(k),n_{y}(k),n_{z}(k)]$ is a unit vector. 
The eigenstate of $\hat{H}_{\rm eff}(k)$ can be represented as $\left|\phi^{s}_{k}\right\rangle=\left|\psi^{s}_{k}\right\rangle\left|k\right\rangle$, where $\left|\psi^{s}_{k}\right\rangle$ is the eigenstate of operator $\textbf{n}(k)\cdot\bm{\sigma}$ and $s=\pm 1$ represents the band index. At the parameter range of $-\pi\leq k \leq \pi$, the eigenenergy of $\hat{H}_{\rm eff}(k)$ forms two symmetrical energy bands with $\pm E_{k}$. In our experiment, this system possesses the chiral symmetry, since $\hat{H}_{\rm eff}(k)$ satisfies $\Gamma\hat{H}_{\rm eff}(k)\Gamma=-\hat{H}_{\rm eff}(k)$ with $\Gamma=\sigma_{z}$ (see section II of SI for more details).

Interestingly, the eigenstates of $\hat{H}_{\rm eff}$ can be directly obtained by measuring the transmission intensities of the cavity. Since the eigenstates of $\hat{H}_{\rm eff}$ form a complete basis, the output state of the cavity could be expanded as $\left|\phi_{out}\right\rangle=\sum_{k,s}T_{k}^{s}\left|\phi_{k}^{s}\right\rangle$. According to the input-output relation of the cavity (see section III of SI for more details), the transmission amplitude can be expressed as 
\begin{equation}
T_{k}^{s}=\frac{|\kappa|^{2}/r}{1-re^{-i(sE_{k}-\beta \Delta L)}}\left\langle\phi_{k}^{s}|\phi_{in}\right\rangle.
\end{equation}
$\Delta L$ denotes the cavity's detuning which equals to $L-2n\pi/\beta$ $(n\in\mathbb{N}^+)$. $\kappa$ and $r$ are the coupling and reflection coefficients respectively of the cavity and they satisfy the condition that $|\kappa|^2+|r|^2=1$, which are nearly same for different OAMs. By choosing an appropriate input state $\left|\phi_{in}\right\rangle$, $\sum_{s}|\left\langle\phi_{k}^{s}|\phi_{in}\right\rangle|$ could be independent on $k$ (see section III of SI for more details).
The whole output transmission intensity, which can be directly measured, is defined as $I_{o}=\sum_{k,s}|T_{k}^{s}|^{2}$. We can find that in Eq. 3, only when the term $\beta \Delta L$ is closest to $sE_{k}$, the photonic output state $\left|\phi_{out}\right\rangle$ is closest to the eigenstate $\left|\phi_{k}^{s}\right\rangle$ and the relevant $I_{o}$ reaches its local maximum at the same time (see section IV of SI for more details). Moreover, the transmission intensity $I_{o}$ contributed by all eigenstate $\left|\phi_{k}^{s}\right\rangle$ of all $k$ corresponds to the density of state (DOS) under renormalization (see section III of SI for more details).

\subsection{Experimental results}
\begin{figure*}[t!]
	\begin{center}
		\includegraphics[width=1.8 \columnwidth]{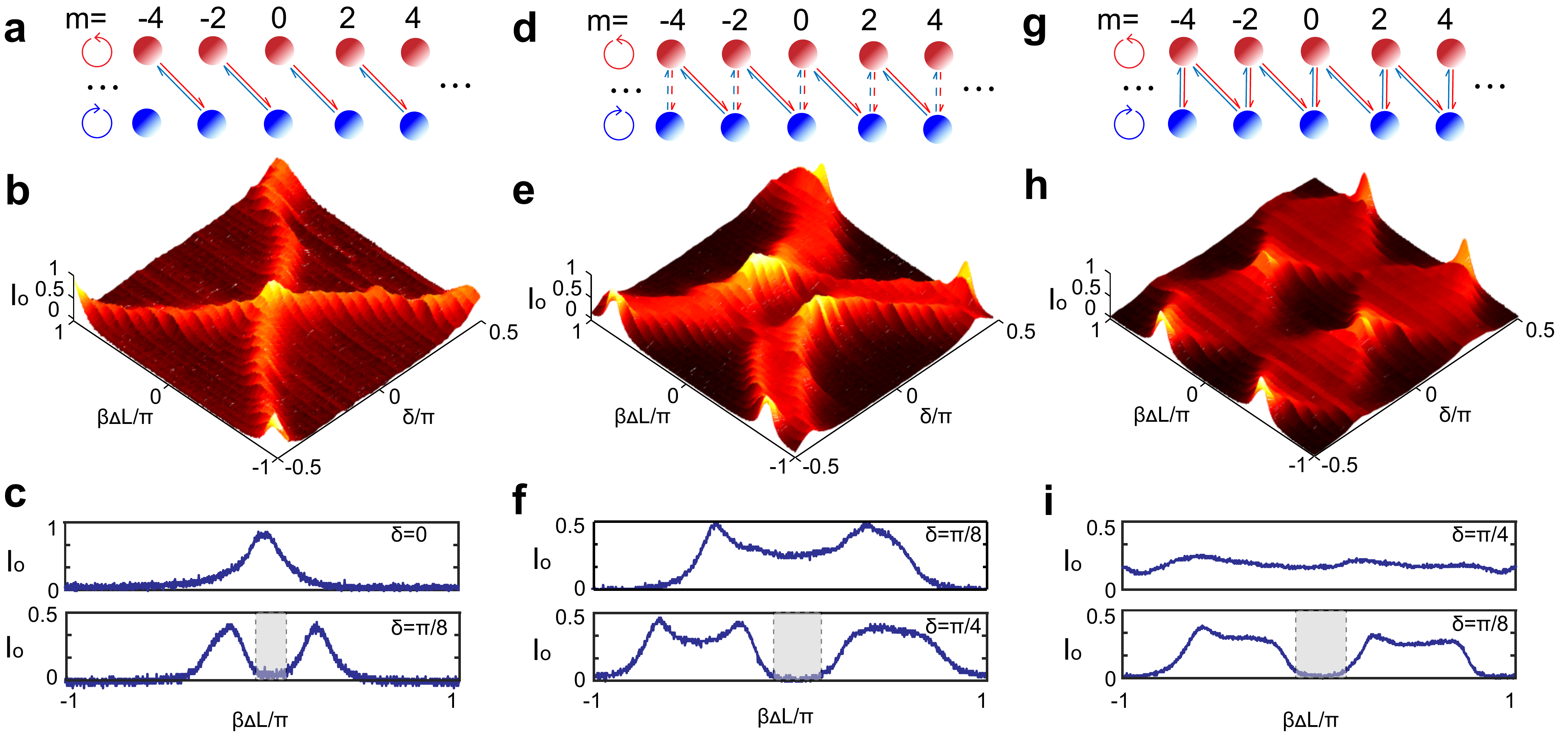}
		\caption{\textbf{detected photonic density of state (DOS) in $(\eta,\delta)$ space}. (\textbf{a, d}, and \textbf{g}). The coupling among the photonic angular momentum with $\eta=0,\pi/8,\pi/4$, respectively. The coupling of SAM modes with different $\eta$ are denoted as no lines, dashed lines and solid lines between nearby modes, respectively. (\textbf{b, e}, and \textbf{h}). The normalised transmission intensity as a function of the normalization cavity detuning parameters  $\beta \Delta L/\pi$ and $\delta/\pi$. (\textbf{c, f}, and \textbf{i}). The spectra of DOS when the gap closes and opens at different $\delta$. The band gap areas are marked in gray. 
		}
		\label{DOS}
	\end{center}
\end{figure*}

From the spectrum of DOS, the energy gap of the system with the zero DOS can also be directly read out. In our experiments, the parameter $\eta$ can be used to control the coupling strength between SAM modes and the $m$-th OAM mode. If $\eta=0$ (there is no WP in the cavity), there is no coupling between $\left|\circlearrowleft, m\right\rangle$ and $\left|\circlearrowright, m\right\rangle$ and the simulated system is reduced to a two-level system. The hopping occurs only between photonic angular momentum states $\left|\circlearrowleft, m\right\rangle$ and $\left|\circlearrowright, m+2\right\rangle$, as shown in FIG. \ref{DOS}a. The measured full spectrum of DOS as a function of $\delta$ and $\beta\Delta L$ is shown in FIG. \ref{DOS}b, and the special cases with $\delta=0$ (closing gap) and $\delta=\pi/8$ (opening gap) are shown in the top and bottom panels of FIG. \ref{DOS}c (the band gap areas are marked in gray), respectively. The closing and opening of energy gap are dependent on the parameter of $\delta$.

With the increase of the parameter $\eta$, the states $\left|\circlearrowleft, m\right\rangle$ and $\left|\circlearrowright, m\right\rangle$ will couple to each other. As a result, the spin-orbital like interaction in the system can be realized with addition coupling between $\left|\circlearrowleft, m\right\rangle$ and $\left|\circlearrowright, m+2\right\rangle$ controlled by $\delta$, as is shown in FIG. \ref{DOS}d. In such kind of situation, the topological phases appear and are protected by the band gap. The closing of the gap indicates the phase transition between the topological phase and the trivial phase. The measured full spectrum of DOS as function of $\delta$ with $\eta=\pi/8$ (the WP is an eighth-wave plate in the cavity) is shown in FIG. \ref{DOS}e. The band gap closes at $\delta=\pm\pi/8$ and $\delta=\pm3\pi/8$ which indicate two phase transitions. In FIG. \ref{DOS}f, the DOS with the gap closing at $\delta=\pi/8$ (top panel) is further compared with the gap opening at $\delta=\pi/4$ (bottom panel). 
The spectrum of DOS as function of $\delta$ with another $\eta=\pi/4$ (the WP is a QWP in cavity) is shown in FIG. \ref{DOS}h. The schematic SOC interation is shown in FIG. \ref{DOS}g. Similarly, there are two gap closure points at $\delta=\pm\pi/4$~\cite{QWe1,QWe2} and the comparation of DOS with the gap closing at $\delta=\pi/4$ (top panel) and with the gap opening at $\delta=\pi/8$ (bottom panel) is shown in  FIG. \ref{DOS}i. 
Worthy to note that our system only has topological protection versus disorder in the coupling constants but not between disorder in the self-energy. The slight deformity of the spectra in Figs. \ref{DOS}c, \ref{DOS}f, and \ref{DOS}i illustrate a distribution of energies around the main energy, which may be due to the imperfect degeneracy of the cavity.

The spectra of DOS display the number of states with the same energy. However, the degeneracy of energy ($E_{-k}=E_{k}$) leads to the indistinguishability of the states with momentum $k$ and $-k$. To determine the relationship between the quasienergy $E_{k}$ and the quasimomemtum $k$, which characterizes the corresponding band structure of the SOC system, we should scan the transmission intensity $I_{k}=\sum_{s}|T_{k}^{s}|^{2}$ as a function of a post-selected Bloch momentum state $\left|k\right\rangle$ (see section VI of SI for more details). 
In experiment, the state projection is carried out by a spatial light modulator (SLM). The state $|k\rangle$ with a superposition of OAM modes is transferred to the Gaussian mode with $m=0$ that is determined by a single mode fiber. However, due to the limitation of the SLM's spatial resolution, we can only project the output state onto $\left|k_{\rm exp}\right\rangle=\sum_{j=-N/2}^{j=N/2}e^{-ijk_{\rm exp}}\left|j\right\rangle$ $(j=m/2)$ with $N$ setting to 12. $\left|k_{\rm exp}\right\rangle$ approaches to $\left|k_{\rm}\right\rangle$ when $N$ increases to infinity. 
The detailed projection process can be found in Methods and the photon distributions after the SLM's modulation are shown in section VII of SI. 

\begin{figure*}[t!]
	\begin{center}
		\includegraphics[width=1.7 \columnwidth]{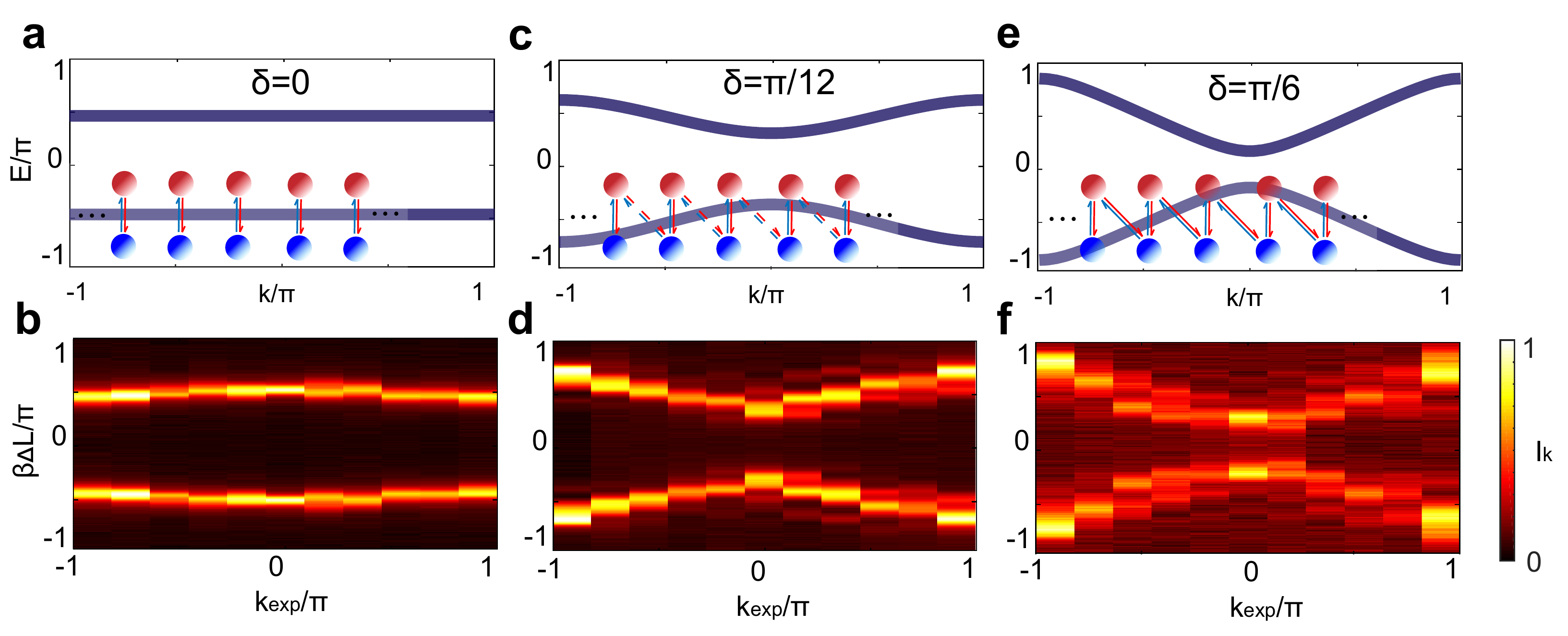}
		\caption{\textbf{Theoretical and experimental band structures.}  (\textbf{a, c}, and \textbf{e}). Theoretically calculated energy band spectra with $\delta=0$, $\pi/12$ and $\pi/6$ when $\eta=\pi/4$, respectively. The x (y)-axis represents the normalised quasimomentum $k/\pi$ (quasienergy $E_{k}/\pi$). The coupling among optical angular momentum are shown in the inserts, in which the increasing strength is denoted as no lines, dashed lines and solid lines between nearby modes, respectively. (\textbf{b, d}, and \textbf{f}). The corresponding experimental energy band of $\delta=0$, $\pi/12$, and $\pi/6$ when $\eta=\pi/4$, respectively. The x (y)-axis represents the normalised cavity detuning $\beta \Delta L/\pi$ (the parameter $k_{\rm exp}/\pi$). }
		\label{band}
	\end{center}
\end{figure*}

The representative theoretical and experimental band structures with different $\delta$ ($0$, $\pi/12$ and $\pi/6$) at $\eta=\pi/4$ are shown in FIG. \ref{band}. Note that due to the limited $N$ ($N=12$), the obtained transmission intensity is a bit concentrated at $k_{\rm exp}=0$ and $\pi$. With the improvement of the spatial resolution of SLM, the experimental results will approach to the ideal results by increasing $N$. Since the band structure represents the refined DOS, the band gap of the simulated topological system also can be read out directly.

It is well known that the SOC systems exhibiting different topological phases can be distinguished by their winding numbers. According to the intrinsic chiral symmetry determined by the form of the unitary operation $\hat{U}(\delta,\eta)$ in one period, the topological bulk invariant in such system can be defined by the winding number of the unit vector $\textbf{n}(k)$ in Hamiltonian $\hat{H}_{\rm eff}$. 
The vector $\textbf{n}(k)$ winds around a fixed axis $z$ with varying $k$ and the trajectory forms a circle on the Bloch sphere. Through performing the polarization Pauli measurements of $\sigma_{i}$ ($i=x,y,z$) on the post-selected state $|k\rangle$, the transmission intensity $I_{k}$ is modified to $I^{i}_{k}=\sum_{s}sn_{i}(k)|T_{k}^{s}|^{2}$. Therefore, the unit vector $\textbf{n}(k)$ and the corresponding winding number can be derived from the variation of the transmission peaks (see section VIII of SI for more details).

\begin{figure*}[t!]
	\begin{center}
		\includegraphics[width=1.7 \columnwidth]{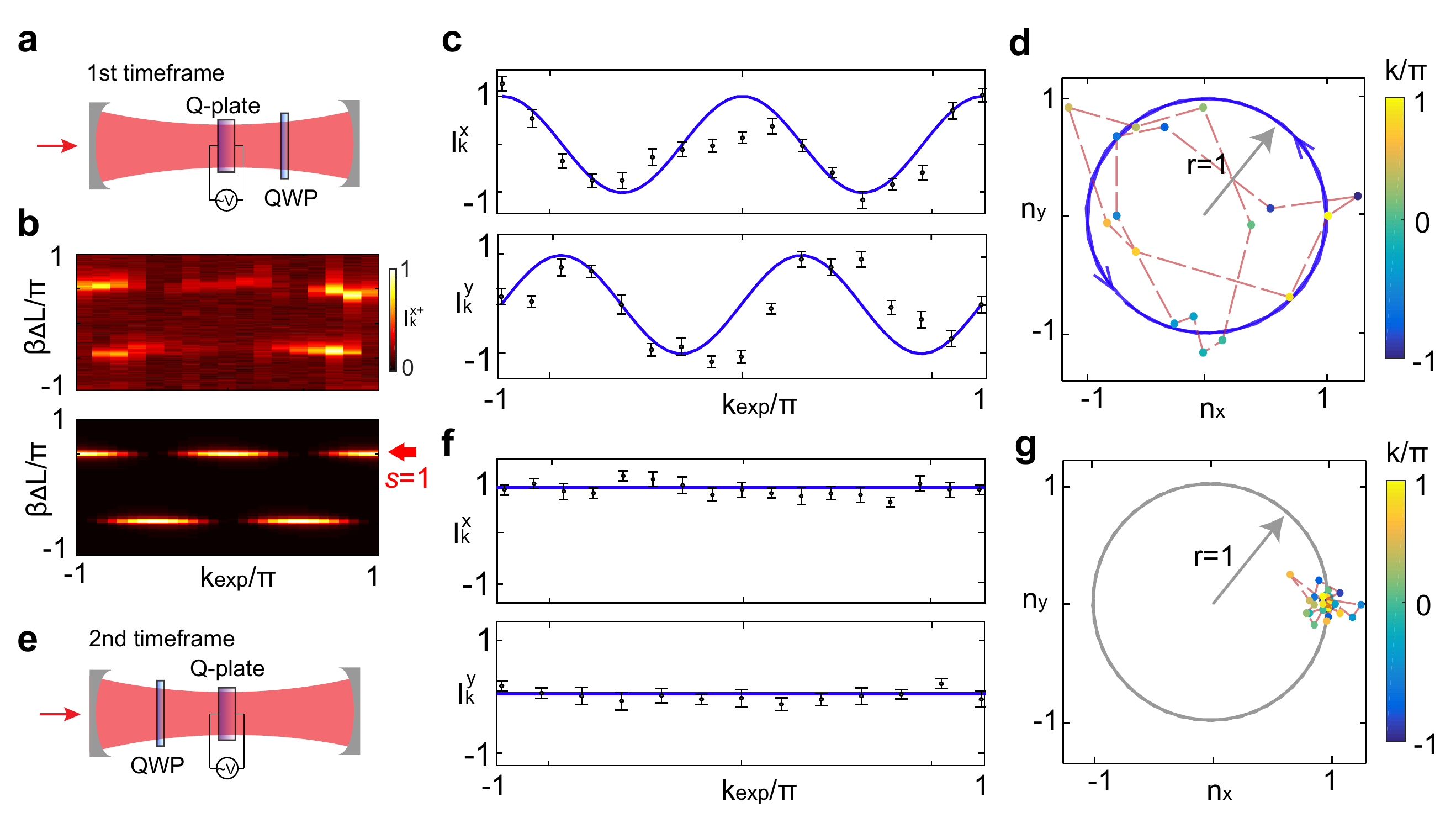}
		\caption{\textbf{Experimental windings.} (\textbf{a} and \textbf{e}). Experimental setups of the 1st and 2nd timeframe settings, respectively. The parameters are set to be $\delta=\pi/2$  for the Q-plate and $\eta=\pi/4$ for the QWP.
		(\textbf{b}). The experimentally measured (top) and numerically simulated (bottom) transmission intensities $I^{x_+}_{k}$ on the horizontal polarization state $(\left|\circlearrowright\right\rangle+\left|\circlearrowleft\right\rangle)/\sqrt{2}$ along the quasimomentum $k_{\rm exp}$. The unit vector $\textbf{n}(k)$ are readout according to $\beta \Delta L>0$ with $s=1$.
		(\textbf{c} and \textbf{f}). The normalised transmitted peaks on basis $\sigma_{x}$ (top) and $\sigma_{y}$ (bottom) of the 1st and 2nd timeframes. The black dots represent the experimental data while the blue curves represent the theoretical predictions. (\textbf{d} and \textbf{g}). The topological windings of measured $(n_{x}, n_{y})$ (color points) at the 1st and 2nd timeframes. The blue circle with the radius of r=1 represent the theoretical predictions.
		}
		\label{wind}
	\end{center}
\end{figure*}

For a periodic driving system, its topological phases should be characterized by two different timeframes (different sequences of the operations in the cavity)~\cite{ABBA1, ABBA2}. The different timeframes give the same dispersion relationship but different windings of the unit vector $\textbf{n}(k)$, which correspond to different topologies (see section II of SI for more details).  For the 1st timeframe in FIG. \ref{wind}a, the evolution operation is $\hat{U}(\delta,\eta)=J_{Q(\delta)}J_{\lambda(\eta)}J_{\lambda(\eta)}J_{Q(\delta)}$. By choosing the parameters to be $\delta=\pi/2$ and $\eta=\pi/4$, the experimental (top panel) and numerical (bottom panel) transmission intensities $I^{x_+}_{k}$ by projecting the SAM mode to the horizontal polarization state $(\left|\circlearrowright\right\rangle+\left|\circlearrowleft\right\rangle)/\sqrt{2}$ representing the eigenstate of $\sigma_{x}$ with eigenvalue $+1$, are shown in FIG. \ref{wind}b.

Different from the transmission intensity $I_{k}$ in FIG. \ref{band}c (which is homogeneous along $k$), the normalized height of the transmitted peaks of $I^{x_+}_{k}$ is periodically modulated along $k$ and exhibit the variations of $sn_{x}(k)$. Furthermore, the height of the transmitted peaks as a function of the quasimomentum $k$ can be devided into two complementary parts: one is for $\beta \Delta L>0$ corresponding to the upper band ($s=1$), and the other is for $\beta \Delta L<0$ corresponding to the lower band ($s=-1$). Since $\boldsymbol{n}(k)$ in each band defines the same winding number, without loss of generality, we choose the upper band to calculate the topological winding number.

We further detect the normalized height of the transmitted peaks of $I^{x_-}_{k}$ by projecting the SAM mode of the output photons to the vertical state $(\left|\circlearrowright\right\rangle-\left|\circlearrowleft\right\rangle)/\sqrt{2}$, which is the eigenstate of $\sigma_{x}$ with eigenvalue $-1$. The value of $n_{x}(k)$ is determined by $I^{x}_{k}=I^{x_+}_{k}-I^{x_-}_{k}$ and the corresponding experimental results are shown in FIG. \ref{wind}c (upper panel). The value of $n_{y}(k)$ can be determined by $I^{y}_{k}=I^{y_+}_{k}-I^{y_-}_{k}$, where $I^{y_+}_{k}$ and $I^{y_-}_{k}$ represent the normalized height of the transmitted peaks by projecting the SAM modes to $(\left|\circlearrowright\right\rangle-i\left|\circlearrowleft\right\rangle)/\sqrt{2}$ and $(\left|\circlearrowright\right\rangle+i\left|\circlearrowleft\right\rangle)/\sqrt{2}$, respectively. The experimental results are shown in the lower panel of FIG. \ref{wind}c. Error bars are estimated according to the fluctuation of the output intensities.

In the $x-y$ plane, we find that the normalized vector [$n_{x}(k),n_{y}(k)$] winds twice anticlockwise around the chiral axis as the quasimomentum $k$ traverses in the first Brillouin zone $[-\pi,\pi]$. The corresponding experimental result is shown in FIG. \ref{wind}d, which indicates that the SOC system possesses a nontrivial topology phase in the 1st timeframe  with $\delta=\pi/2$ and $\eta=\pi/4$. The nontrivial topological insulator would support edge states at interfaces where the topological invariant changes. For instance, when the coupling between SAM breaks at the centre of the lattice ($m=0$) with some unique designs (see section V of SI for more details), the interface between the nontrivial topological bulk and ``vacuum" can support edge states. It is worth mentioning that the winding numbers are protected by symmetry when the strength of the disorder is less than the bandgap. However, the increasing disorder can move the edge to the bulk bands.

On the other hand, the 2nd timeframe, with the evolution unitary $\hat{U}^{'}(\delta,\eta)=J_{\lambda(\eta)}J_{Q(\delta)}J_{Q(\delta)}J_{\lambda(\eta)}$, is constructed by exchanging the Q-plate and WP in the cavity, which is shown in FIG. \ref{wind}e. The corresponding experimental results of $n_{x}(k)$ and $n_{y}(k)$ are shown in FIG. \ref{wind}f. In the $x-y$ plane, $\textbf{n}(k)$ winds 0 round in FIG. \ref{wind}g. As a result, although the SOC system in both timeframes have the same band structure, they have completely different winding numbers. 

\section{Discussion}
In conclusion, we have experimentally demonstrated a compact optical spin-orbital coupling system in a degenerate cavity. The optical OAM degree of freedom serves as a synthetic dimension, and the interaction strength of SOC, introduced by the Q-plate and WP in the cavity, can be tuned conveniently. The DOS, band structures, and topological windings of the synthetic topological insulator, which shares the famous features of SSH model, are directly obtained by detecting the transmission intensity of the cavity. Through manipulating the parameters of the cavity, we obtain multiple DOS to observe the closing of the band gap directly and investigate the topological phases in different time frames.

Our work provides a versatile platform based on an OAM degenerate cavity to explore richer topological physics. Higher-dimensional physics can then be exploited in the compact platform. The 2-leg ladder model can be achieved by introducing an additional Q-plate with $q=-1$ into the cavity. Moreover, the setup is compatible with other synthetic dimensions, including the frequency degree of freedom~\cite{Fan3}. By introducing the external gauge fields in the cavity, topological systems with the famous Hofstadter's butterfly spectrum can be directly investigated~\cite{Luo1}. Two-dimensional topological systems are generally more robust than one-dimensional topological systems. The topological properties of one-dimensional systems come purely from symmetries, while the topological properties of two-dimensional systems come from gauge fields.

Moreover, non-Hermitian interactions would be realized through involving the  gain/loss of the spin degree of freedom in the cavity and the non-Hermitian physics can also be well-studied~\cite{PPBS1,PPBS2}. The introduced nonlinearities or gain/loss would create "boundaries" inside the bulk and make the topological systems without boundaries still present topological bulk properties.
The degenerate cavity containing many optical angular momentum may also suit for employing as all-optical devices, such as quantum memory and optical filters~\cite{Luo2}.

\section{Methods}
\subsection{The operation of Q-plate}
The Q-plate is composed of liquid crystal molecules with different optical axes, each of which is equivalent to a half-wave plate \cite{Qp}. The optical axis of cylindrical coordinate satisfies 
\begin{equation}
\alpha(r, \phi)=q\phi,
\end{equation}
where $q$ are constants. The Jones formalism of Q-plate $\textbf{M}_{Q}$ can be written as
\begin{equation}
\textbf{M}_{Q}=
%[J_{0}(\beta)+J_{1}(\beta)e^{i(k_{y}+\varphi)}-J_{1}(\beta)e^{-i(k_{y}+\varphi)}]
\cos(\delta/2)
\textbf{I}
+
i\sin(\delta/2)
\left[
\begin{array}{cc}
\cos 2\alpha & \sin 2\alpha\\
\sin 2\alpha & -\cos 2\alpha
\end{array}
\right],
\end{equation}
where $\delta$ is the optical retardation and is controlled by the applied electric field. Within the paraxial approximation, a left (right)-circular polarized plane wave $\textbf{E}=E_{0}\left[
\begin{array}{c}
1 \\
\pm i
\end{array}
\right]e^{im\phi}$, denoted as $\left|\circlearrowleft(\circlearrowright), m\right\rangle$ (where $\circlearrowleft (\circlearrowright)$ denotes the left (right)-circular polarized SAM modes and $m$ is the topological charge of OAM, passes through the Q-plate, and the plane wave would change to
\begin{equation}
\frac{\textbf{E}^{'}}{E_{0}}%=\textbf{M}_{Q}E_{0}
% \left[
% \begin{array}{c}
% 1 \\
% \pm i
% \end{array}
% \right]
=\cos(\delta/2)
\left[
\begin{array}{c}
1 \\
\pm i
\end{array}
\right]
+i\sin(\delta/2)e^{i[(\pm 2q+m)\phi]}
\left[
\begin{array}{c}
1 \\
\mp i
\end{array}
\right].
\label{Qpo}
\end{equation}
The vortex phase $e^{\pm i2q\phi}$ (topological charge is $2q$) is introduced during the spin-to-orbital angular momentum conversion.\\

\subsection{The phase hologram}
Within the paraxial approximation, the state $\left|m\right\rangle$ of the photon carrying OAM with topological charge $m$ can be approximately expressed as
\begin{equation}
\left|m\right\rangle=E_{0}e^{im\phi},
\end{equation}
where the phase $\phi=\tan^{-1}(y/x)$ in cartesian coordinates $(x,y)$. The phase of $\left|k_{\rm exp}\right\rangle$ at the position of ($x,y$) is
\begin{equation}
\begin{aligned}
\varphi_{(k_{\rm exp},N)}(x,y)&=\arg\sum_{j=-N/2}^{j=N/2}e^{-ijk_{\rm exp}}\left|j\right\rangle\\
&=\arg\sum_{m=-N}^{m=N}e^{-im(k_{\rm exp}/2+\phi)},\\
%&=&\arg\delta(k_{\rm exp}, \phi).
\end{aligned}
\end{equation}
where $j=m/2$.
For an 8-bit SLM, the modulation phase from 0 to $2\pi$ is mapped to gray value 0 to 255. The hologram of basis $\left|k_{\rm exp}\right\rangle\langle k_{\rm exp}|$ is given by
\begin{equation}
H_{(k_{\rm exp},N)}(x,y)=[{\rm mod}(\varphi_{(k_{\rm exp},N)}(x,y), 2\pi)\times 255],
\end{equation}
where $H_{(k_{\rm exp},N)}(x,y)$ represents the gray value at ($x,y$) position on SLM.
The hologram $H_{(k_{\rm exp},N)}(x,y)$ on SLM is parameterized by $(k_{\rm exp},N)$.

\section{Data Availability}
The data that support the findings of this study are available from the corresponding author on reasonable request.

\bibliographystyle{unsrt}

\clearpage
%%%%%%%%%%%%%%%%%%%%%%%%%%%%%%%%%%%%%%%%%%%%%%%%%%%%%%%%%%%%%%%%%%%%%%%%%%%%%%%%%%%%%%%%%%%%%%%%%%%%%%%%%%%%%%%%%%%%%%%%%%%%%%%%%
\section{Acknowledgement}
We acknowledge the discussion with Ze-Di Cheng. This work was supported by the National Key Research and Development Program of China (Grant No. 2016YFA0302700), the National Natural Science Foundation of China (Grants No. 11874343, 61725504, 61327901, 61490711, 11774335, 11821404 and U19A2075), the Key Research Program of Frontier Sciences, Chinese Academy of Sciences (CAS) (Grant No. QYZDY-SSW-SLH003), Science Foundation of the CAS (No. ZDRW-XH-2019-1), Anhui Initiative in Quantum Information Technologies (AHY060300 and AHY020100), the Fundamental Research Funds for the Central Universities (Grant No. WK2030380017 and WK2470000026).\\

\section{Author contributions}
M. Y. and Y.-W. L. experimented with the assistant of J.-S. X. and Z. H. L.;
M. Y, H.-Q. Z. and Y.-J. H. contributed to the theoretical analysis with the help of Z. W. Z. and X. X. Z;
J.-S. X, Y.-J. H, C.-F. L. and G.-C. G. supervised the project.
All authors read the paper and discussed the results.\\

\section{Additional Information}
Competing interests: The authors declare no competing financial interests.

\end{document}

\title{Supplementary Information for: Topological band structure via twisted photons in a degenerate cavity}
%%%%%%%%%%%%%%%%%%%%%%%%%%%%%%%%%%%%%%%%%%%%%%%%%%%%%%%%%%%%%%%%%%%

\author{Mu Yang\red{\footnote{These authors contribute equally to this work\label{Contribute}}}}
\affiliation{CAS Key Laboratory of Quantum Information, University of Science and Technology of China, Hefei 230026, People's Republic of China}
\affiliation{CAS Center For Excellence in Quantum Information and Quantum Physics, University of Science and Technology of China, Hefei 230026, People's Republic of China}

\author{Hao-Qing Zhang\textsuperscript{\ref {Contribute}}}
\affiliation{CAS Key Laboratory of Quantum Information, University of Science and Technology of China, Hefei 230026, People's Republic of China}
\affiliation{CAS Center For Excellence in Quantum Information and Quantum Physics, University of Science and Technology of China, Hefei 230026, People's Republic of China}

\author{Yu-Wei Liao\textsuperscript{\ref {Contribute}}}
\affiliation{CAS Key Laboratory of Quantum Information, University of Science and Technology of China, Hefei 230026, People's Republic of China}
\affiliation{CAS Center For Excellence in Quantum Information and Quantum Physics, University of Science and Technology of China, Hefei 230026, People's Republic of China}

\author{Zheng-Hao Liu}
\affiliation{CAS Key Laboratory of Quantum Information, University of Science and Technology of China, Hefei 230026, People's Republic of China}
\affiliation{CAS Center For Excellence in Quantum Information and Quantum Physics, University of Science and Technology of China, Hefei 230026, People's Republic of China}

\author{Zheng-Wei Zhou}
\affiliation{CAS Key Laboratory of Quantum Information, University of Science and Technology of China, Hefei 230026, People's Republic of China}
\affiliation{CAS Center For Excellence in Quantum Information and Quantum Physics, University of Science and Technology of China, Hefei 230026, People's Republic of China}

\author{Xing-Xiang Zhou}
\affiliation{CAS Key Laboratory of Quantum Information, University of Science and Technology of China, Hefei 230026, People's Republic of China}
\affiliation{CAS Center For Excellence in Quantum Information and Quantum Physics, University of Science and Technology of China, Hefei 230026, People's Republic of China}

\author{Jin-Shi Xu}\email{jsxu@ustc.edu.cn}
\affiliation{CAS Key Laboratory of Quantum Information, University of Science and Technology of China, Hefei 230026, People's Republic of China}
\affiliation{CAS Center For Excellence in Quantum Information and Quantum Physics, University of Science and Technology of China, Hefei 230026, People's Republic of China}

\author{Yong-Jian Han}\email{smhan@ustc.edu.cn}
\affiliation{CAS Key Laboratory of Quantum Information, University of Science and Technology of China, Hefei 230026, People's Republic of China}
\affiliation{CAS Center For Excellence in Quantum Information and Quantum Physics, University of Science and Technology of China, Hefei 230026, People's Republic of China}
	
\author{Chuan-Feng Li}\email{cfli@ustc.edu.cn}
\affiliation{CAS Key Laboratory of Quantum Information, University of Science and Technology of China, Hefei 230026, People's Republic of China}
\affiliation{CAS Center For Excellence in Quantum Information and Quantum Physics, University of Science and Technology of China, Hefei 230026, People's Republic of China}

\author{Guang-Can Guo}
\affiliation{CAS Key Laboratory of Quantum Information, University of Science and Technology of China, Hefei 230026, People's Republic of China}
\affiliation{CAS Center For Excellence in Quantum Information and Quantum Physics, University of Science and Technology of China, Hefei 230026, People's Republic of China}
\onecolumngrid
\setcounter{equation}{0}
\setcounter{figure}{0}
\renewcommand{\theequation}{S\arabic{equation}}
\renewcommand{\thefigure}{S\arabic{figure}}

\maketitle	

\tableofcontents

\section{Details of the experimental setup} 
\begin{figure*}[t!]
	\begin{center}
		\includegraphics[width=0.8 \columnwidth]{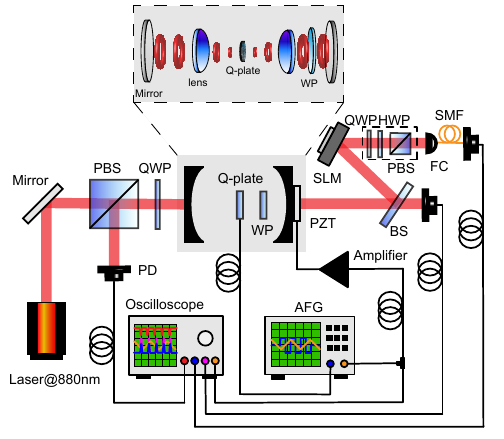}
		\caption{ \textbf{Experimental setup.} The cavity of 0.4 m length is pumped by an infrared laser beam at 880 nm in a Gaussian mode. A quarter-wave plate (QWP) is used to rotate the polarization of the incident beam. A photodetector (PD) in the reflection path of the polarization beam splitter (PBS) is used to detect the reflected optical intensity from the cavity. The transmitted signal is divided into two paths by a beam splitter (BS). One of which is detected directly and the other is modulated firstly by a spatial light modulator (SLM) and then coupled to a single-mode fiber (SMF). A QWP, a half-wave plate (HWP), and a PBS in the dotted box before the fiber coupler (FC) is used for the post-selection of polarization. An arbitrary function generator (AFG) modulates the Q-plate and periodically drives the Piezoelectric Transducer (PZT) attached to the output couple BS of the cavity after amplification to change the length of the cavity. All electrical signals are recorded by the oscilloscope. The inset shows the details of the cavity which consists of two Mirrors and two lenses. The Q-plate and a wave-plate (WP) in the cavity are used to modulate the photon modes, which are represented as red circles. 
		}
		\label{setup}
	\end{center}
\end{figure*}

 We use the device shown in FIG. \ref{setup} to investigate the topological properties of the spin-orbit coupling system. We use an infrared continuous wave (CW) laser with the Gaussian mode at $\lambda=880$ nm. The polarization of the laser is prepared to be left or right circular ($\circlearrowleft$ or $\circlearrowright$) after passing through the polarization beam splitter (PBS) and a quarter-wave plate (QWP) with the optical-axis setting at 45$^\circ$. The photons are coupled into the cavity by the first mirror with a ratio between transmission and reflection of 5/95. The photons that are not coupled into the cavity are detected by a photodetector (PD). Since the constructed degenerate optical resonant cavity supports all the Laguerre-Gauss (LG) modes that form a complete basis, the input laser mode does not need to be specially adjusted.

The degenerate optical cavity has been theoretically investigated \cite{cavity1,cavity2}. In experiment, the degenerate cavity consist of two plane mirrors and two lenses of focal length $f=0.1$ m. The free spectral range (FSR) of the cavity is about 375 MHz, while the linewidth is about 13.6 MHz. A Q-plate with $q=1$ is placed in the center of the cavity, on which the electrostatic field is controlled by an arbitrary function generator (AFG). A ($\eta/\pi$)-wave plate (WP) is set behind to rotate the polarization (e.g., $\eta=\pi/4$ for the quarter-wave plate). To scan the cavity length $\Delta L$, a piezoelectric transducer (PZT) is pasted on the second mirror and is driven by an amplified periodic triangular wave signal generated by the AFG. The scanning frequency of the triangular wave signal is set at 40 Hz such that the system reaches a steady state at each frequency. To reduce the cavity's dissipation ($\alpha$), all the optical elements in the cavity are coated with anti-reflection films. The linewidth of the cavity can be reduced to obtain sharper transmission peaks. 

The photons are out-coupled by the second mirror with the ratio of  1/99. The output photons are separated into two paths by a beam splitter (BS). The transmitted photons are detected directly by a PD, while the reflected parts are first modulated by a spatial light modulator (SLM) and post-selected by a single mode fiber (SMF). When detecting topological windings, a QWP, a half-wave plate (HWP), and a PBS in the dashed panel are set before the fiber coupler (FC) for the post-selection of polarization. All the signals detected by the PDs are recorded in an oscilloscope with a 1 GHz bandwidth, which allows readind the system's eigenenergy directly.

In our experiment, the length of the synthetic dimension is limited by the size of the minimum aperture of the Q-plate. With the increasing of topological charge number $m$ of orbital angular momentum (OAM)  modes, the transverse radius of maximum field amplitude $r_{m}$ will increase as $r_{m}=\omega_{0}\sqrt{m/2}$~\cite{R}. $\omega_{0}$ represents the waist radius of Gaussian mode ($m=0$). In this experiment, the radius of the Q-plate (minimal aperture) is 0.25 mm and $\omega_{0}$ is about 80 $\mu$m. The maximal topological charge number $m$ is $2r_{m}^2/\omega_{0}^2\approx1.95\times 10^3$ . The cavity supports the even order OAM modes ranging $-1.95\times 10^3<m<1.95\times 10^3$ and the length of the synthetic dimension is about $1.95\times 10^3$. % which is much larger than the previous simulation systems.

On the other hand, the cavity loss determines the average  lifetime of photons. The total loss $\alpha$ of the cavity is about $0.1$. The average lifetime of photons becomes $\tau=L/(\alpha c)=10L/c$ with $L$ and $c$ representing the length of the cavity and the speed of the photons, respectively. Therefore, the photons can pass through the Q-plate 10 times on average. The average charges of degenerate orbital angular momenta inside the degenerate cavity is located in $m\in[-20, 20]$. The lifetime can significantly increase by introducing an optical amplifier into the cavity.

\section{Dispersion relation of the cavity}
In this section, we derive the dispersion relation for the cavity without input and output. As the photons propagate in a steady cavity, the state $\left|\phi(t)\right\rangle=(...,\phi(t)_{\circlearrowleft,m-1},\phi(t)_{\circlearrowright,m-1},\phi(t)_{\circlearrowleft,m}, \phi(t)_{\circlearrowright, m},...)$ at time $t$ must satisfy the condition of mode self-reproduction, denoted as
\begin{equation}
\begin{aligned}
\left|\phi(t+T)\right\rangle=e^{-i\beta L}\left|\phi(t)\right\rangle,
\label{ba0}
\end{aligned}
\end{equation}
where $\beta=\Omega/c+i\alpha$ and $L$ is one round trip
(one period) length of the cavity.  $\Omega$ is the resonant frequency of the vacuum cavity and $\alpha$ is the attenuation coefficient. $T=L/c$, where $c$ represnts the speed of light.
On the other hand, according from the method, the action of Q-plate ($q=1$) is described as
\begin{equation}
\begin{aligned}
J_{Q(\delta)}=\sum_{m}\cos(\delta/2)(a_{\circlearrowleft,m}^{\dag}a_{\circlearrowleft,m}+a_{\circlearrowright,m}^{\dag}a_{\circlearrowright,m})+i\sin(\delta/2)(a_{\circlearrowright,m+2q}^{\dag}a_{\circlearrowleft,m}+\mathrm{h.c.}).
\end{aligned}
\end{equation}
The action of $\eta/\pi$-wave plate (WP) can be described as
\begin{equation}
\begin{aligned}
J_{\lambda(\eta)}=\sum_{m}\cos(\eta)(a_{\circlearrowleft,m}^{\dag}a_{\circlearrowleft,m}+a_{\circlearrowright,m}^{\dag}a_{\circlearrowright,m})+i\sin(\eta)(a_{\circlearrowright,m}^{\dag}a_{\circlearrowleft,m}+\mathrm{h.c.}).
\end{aligned}
\end{equation}
When photons pass a round trip through the Q-plate and $\eta/\pi$-wave plate, the operation on the photons in the cavity can be described as 
\begin{equation}
\begin{aligned}
\hat{U}=J_{Q(\delta)}J_{\lambda(\eta)}J_{\lambda(\eta)}J_{Q(\delta)}.
\label{ab}
\end{aligned}
\end{equation}
When the evolutionary period is guaranteed, the photon state in the cavity satisfies
\begin{equation}
\begin{aligned}
\left|\phi(t+T)\right\rangle=\hat{U}\left|\phi(t)\right\rangle.
\label{ba00}
\end{aligned}
\end{equation}
According to Eq. \ref{ba0} and \ref{ba00}, we find
\begin{equation}
e^{-i\beta L}\left|\phi(t)\right\rangle=\hat{U}\left|\phi(t)\right\rangle.
\label{Ua10}
\end{equation}
Here we define an effective Hamiltonian of $\hat{H}_{\rm eff}=i\log\hat{U}$. In the quasi-momentum space, we define the Bloch modes $\left|k\right\rangle=\sum_{j}e^{-ijk}\left|j\right\rangle$ $(j=m/2)$. The Hamiltonian is denoted as $\hat{H}_{\rm eff}(k)=\textbf{n}(k)\cdot\boldsymbol{\sigma}E_{k}\left|k\right\rangle\left\langle k\right|$, where $\textbf{n}(k)=[n_{x}(k),n_{y}(k),n_{z}(k)]$ is an unit vector, and $\boldsymbol{\sigma}=[\sigma_{x},\sigma_{y},\sigma_{z}]$ is the Pauli matrix.
The eigenstate of $\hat{H}_{\rm eff}(k)$ can be written as $\left|\phi^{s}_{k}\right\rangle=\left|\psi^{s}_{k}\right\rangle\left|k\right\rangle$, where $\left|\psi^{s}_{k}\right\rangle=(\psi_{\circlearrowleft,k}^{s},\psi_{\circlearrowright,k}^{s})^{T}$ and
Eq. \ref{Ua10} becomes
\begin{equation}
e^{-i\beta L}\left|\psi^{s}_{k}\right\rangle=\hat{U}_{k}\left|\psi^{s}_{k}\right\rangle,
\label{Ua2}
\end{equation}
where $s=\pm1$ related to the SAM denotes the upper and lower energy bands. The unitary evolution in the momentum space becomes $\hat{U}_{k}=J_{Q_{k}(\delta)}J_{\lambda_{k}(\eta)}J_{\lambda_{k}(\eta)}J_{Q_{k}(\delta)}$. $J_{Q_{k}(\delta)}$ and $J_{\lambda_{k}(\eta)}$ represent the operations of Q-plate and WP in reciprocal space, respectively, which are given by
\begin{equation}
J_{Q_{k}(\delta)}=\left[
\begin{array}{cc}
\cos(\delta/2)  & i\sin(\delta/2)e^{-ik}  \\
i\sin(\delta/2)e^{ik} & \cos(\delta/2)
\end{array}
\right],
\label{Qk}
\end{equation}
and
\begin{equation}
J_{\lambda_{k}(\eta)}=\left[
\begin{array}{cc}
\cos(\eta)  & i\sin(\eta)  \\
i\sin(\eta) & \cos(\eta)
\end{array}
\right].
\label{Qk}
\end{equation}
The Eq. \ref{Ua2} can be rewritten as
\begin{equation}
e^{-i\beta L}\left|\psi^{s}_{k}\right\rangle=\hat{U}_{k}\left|\psi^{s}_{k}\right\rangle=e^{-i\hat{H}_{\rm eff}(k)}\left|\psi^{s}_{k}\right\rangle.
\label{Ua3}
\end{equation}
The evolution satisfies $\hat{U}_{k}=\cos(E_{k})I+i\sin(E_{k})\textbf{n}(k)\cdot\boldsymbol{\sigma}$. Compared with the Eq. \ref{Ua3}, we can obtain
\begin{equation}
\begin{array}{cc}
s E_{k}(\eta,\delta)=s\cos^{-1}[\sin2\eta\cos k\sin\delta-\cos2\eta\cos\delta],\\
sn_{x}=[\cos 2k \sin^{2}\dfrac{\delta}{2}\sin 2\eta-\cos k \cos 2\eta \sin \delta- \cos^{2}\dfrac{\delta}{2}\sin 2\eta]/\sin s E_{k},\\
sn_{y}=[\sin 2k \sin^{2}\dfrac{\delta}{2}\sin 2\eta-\sin k \cos 2\eta \sin \delta]/\sin s E_{k},\\
sn_{z}=0,
%\beta L=-E_{k}(\eta,\delta).
\end{array}
\label{dispersion}
\end{equation}
where $E_{k}$ is the energy dispersion relation of the cavity. 
The unit vector $\textbf{n}(k)=[n_{x}(k),n_{y}(k),n_{z}(k)]$ reveals the topological winding numbers of the system as discussed later. We can observe the windings of the unit vector $\textbf{n}(k)$ and $-\textbf{n}(k)$ for the upper and lower bands. Moreover, we can find $\hat{H}_{\rm eff}(k)$ meets $\Gamma\hat{H}_{\rm eff}(k)\Gamma=-\hat{H}_{\rm eff}(k)$ with $\Gamma=\sigma_{z}$, which means the system has chiral symmetry. 

The timeframes are the
time evolution with different starting points, which are unique properties in the periodically
driven system. If the photons pass a round through $\eta/\pi$-WP firstly and then the Q-plate, denoted as the second time frame, the evolution operator $\hat{U}^{'}_{k}$ can be rewritten as
\begin{equation}
\begin{array}{rcl}
\hat{U}^{'}_{k}&=&J_{\lambda_{k}(\eta)}J_{Q_{k}(\delta)}J_{Q_{k}(\delta)}J_{\lambda_{k}(\eta)}.
\end{array}
\end{equation}
Similarly, the energy dispersion and unit victors of the second timeframe are given by
\begin{equation}
\begin{array}{cc}
s E^{'}_{k}(\eta,\delta)=s\cos^{-1}[\sin2\eta\cos k\sin\delta-\cos2\eta\cos\delta],\\
sn^{'}_{x}=-[\cos k \cos 2\eta \sin \delta +\cos \delta \sin 2\eta]/\sin s E_{k},\\
sn^{'}_{y}=-[\sin k \sin \delta]/\sin s E_{k},\\
sn^{'}_{z}=0.
%\beta L=-E_{k}(\eta,\delta).
\end{array}
\end{equation}
Obviously, these two time frames have the same energy dispersion relation but the different three-dimensional unit vector $\textbf{n}(k)$. The winding number of the unit vector $n(k)$ is the topological invariant, protected by the chiral symmetry. Therefore, the two timeframes have different topological invariants and correspond to different topologies. 

\section{Direct measurement of the density of states} 
In this section, we turn to an open system and demonstrate the method to directly measure the density of states (DOS) from the cavity output. 
The coupling of the cavity mirror can be described by \cite{coupling1,coupling2}
\begin{equation}
\left[
\begin{array}{c}
   \phi_{out} \\
   a 
\end{array}
\right]_{\circlearrowleft(\circlearrowright),m}
=
\left[
\begin{array}{cc}
    r & \kappa  \\
    -\kappa^{*} & r^{*}
\end{array}
\right]
\left[
\begin{array}{c}
    \phi_{in} \\ 
    b
\end{array}
\right]_{\circlearrowleft(\circlearrowright),m},
\label{cd}
\end{equation}
where $\kappa=i|\kappa|$ and $r=|r|$.
$m$ represents the OAM topological charge. 
$\phi_{in}$ ($\phi_{out}$) represents the input (output) photonic state and $a$ ($b$) represents the state before (after) modulation in the cavity.
Consider phase accumulations as the photons propagate around the cavity, the photonic amplitude $a_{\circlearrowleft(\circlearrowright),m}$ should be rewritten as
\begin{equation}
\begin{aligned}
a_{\circlearrowleft(\circlearrowright),m}\to e^{-i\beta L}a_{\circlearrowleft(\circlearrowright),m}.
\label{ba}
\end{aligned}
\end{equation}
% Combining Eq. \ref{cd} and \ref{ba}, we obtain

Combining Eq. \ref{cd} and \ref{ba}, we find
\begin{equation}
b_{\circlearrowleft(\circlearrowright),m} =\frac{1}{r^{*}}(e^{-i\beta L}a_{\circlearrowleft(\circlearrowright),m}+\kappa^{*}\phi_{in,\circlearrowleft(\circlearrowright),m}),
\label{Sb}
\end{equation}
and
\begin{equation}
\phi_{out,\circlearrowleft(\circlearrowright),m}  =\frac{1}{r^{*}} (\kappa e^{-i\beta L}a_{\circlearrowleft(\circlearrowright),m}+\phi_{in,\circlearrowleft(\circlearrowright),m}).
\label{Ob}
\end{equation}
Representing the states as the state vectors, Eq. \ref{Sb} and Eq. \ref{Ob} become
\begin{equation}
\left|b\right\rangle =\frac{1}{r^{*}}(e^{-i\beta L}\left|a\right\rangle+\kappa^{*}\left|\phi_{in}\right\rangle),
\label{Sb1}
\end{equation}
and 
\begin{equation}
\left|\phi_{out}\right\rangle =\frac{1}{r^{*}} (\kappa e^{-i\beta L}\left|a\right\rangle+\left|\phi_{in}\right\rangle),
\label{Ob1}
\end{equation}
where  $\left|\phi_{in}\right\rangle=(...,\phi_{in,\circlearrowleft,m-1},\phi_{in,\circlearrowright,m-1},\phi_{in,\circlearrowleft,m}, \phi_{in,\circlearrowright, m},...)$, so are $\left|a\right\rangle$, $\left|b\right\rangle$ and $\left|\phi_{out}\right\rangle$.

By taking $\left|b\right\rangle=\hat{U}\left|a\right\rangle$ into Eq. \ref{Sb1}, we can get
\begin{equation}
\frac{1}{r^{*}}(e^{-i\beta L}\left|a\right\rangle_{n}+\kappa^{*}\left|\phi_{in}\right\rangle)=\hat{U}\left|a\right\rangle_{n-1},
\label{sa1}
\end{equation}
where $n$ represents the loop number of the photons running in the cavity. Note that $\left|a\right\rangle_{n-1}=\left|a\right\rangle_{n}$ if $n\to\infty$. Initially, there is no photon in the cavity, which means $\left|a\right\rangle_{0}=\textbf{0}$. After $n$ loop number, we get 
\begin{equation}
\left|a\right\rangle_{n} = -\kappa^{*}e^{i\beta L}\sum_{n}(r^{*})^{n}e^{in\beta L}\hat{U}^{n}\left|\phi_{in}\right\rangle.
\label{an}
\end{equation}

Combining the Eq. \ref{Ob1} and \ref{an}, we can get the output state as
\begin{equation}
\left|\phi_{out}\right\rangle=\frac{1}{r^{*}}\left|\phi_{in}\right\rangle-\frac{|\kappa|^{2}}{r^{*}}\sum_{n}(r^{*})^{n}e^{in\beta L}\hat{U}^{n}\left|\phi_{in}\right\rangle.
\label{out1}
\end{equation}
The first term on the right-hand side represents the direct reflection of $\left|\phi_{in}\right\rangle$. The second term represents the transmission of the field, and we redefine the $\left|\phi_{out}\right\rangle$ as
\begin{equation}
\left|\phi_{out}\right\rangle=-\frac{|\kappa|^{2}}{r^{*}}\sum_{n}(r^{*})^{n}e^{in\beta L}\hat{U}^{n}\left|\phi_{in}\right\rangle.
\end{equation}
The eigenstates $\left|\phi_{k}^{s}\right\rangle=\left|\psi_{k}^{s}\right\rangle\left|k\right\rangle$ of the Hamiltonian $\hat{H}_{\rm eff}(k)$ form a set of complete basis for expanding $\left|\phi_{out}\right\rangle$. We set the input field to be $\left|\phi_{in}\right\rangle=\left|\phi_{in}^{s}\right\rangle\left|m_{0}\right\rangle$ with $\left|m_{0}\right\rangle$ representing a special momentum state.
The transmission field $\left|\phi_{out}\right\rangle$ can be written as 
\begin{equation}
\begin{array}{rcl}
\left|\phi_{out}\right\rangle&=&-\frac{|\kappa|^{2}}{r^{*}}\sum_{k,s}\sum_{n\to\infty}(r^{*})^{n}e^{in\beta L}e^{-inE_{k}} \langle k\left|m_{0}\right\rangle \langle\psi_{k}^{s}\left|\phi_{in}^{s}\right\rangle\left|\psi_{k}^{s}\right\rangle\left|k\right\rangle\\
&=&\sum_{k,s}\frac{-|\kappa|^{2}/r^{*}}{1-r^{*}e^{-i(sE_{k}-\beta L)}}\langle k\left|m_{0}\right\rangle \langle\psi_{k}^{s}\left|\phi_{in}^{s}\right\rangle\left|\psi_{k}^{s}\right\rangle\left|k\right\rangle,
\end{array}
\end{equation}
where $s E_{k}$ represents the eigenenergy of the Hamiltonian $\hat{H}_{\rm eff}$.
Taking $\Delta L$= $L-2n\pi/\beta$  ($n\in\mathbb{N}^+$ and $\beta \Delta L < 2\pi$), we define the transmission coefficient $T_{k}^{s}$ as
\begin{equation}
T_{k}^{s}=\frac{-|\kappa|^{2}/r}{1-re^{-i(sE_{k}-\beta \Delta L)}}\langle k\left|m_{0}\right\rangle \langle\psi_{k}^s\left|\phi_{in}^{s}\right\rangle.
\end{equation}
The intensity $I_{o}=|\phi_{out}|^{2}$ of the output field is 
\begin{equation}
\begin{array}{rcl}
I_{o}&=&\sum_{s}\sum_{kk^{'}}\langle k^{'}|\langle\psi_{k^{'}}^{s}|(T_{k^{'}}^{s})^{*}T_{k}^{s}\left|\psi_{k}^{s}\right\rangle\left|k\right\rangle\\
&=&\sum_{s}\sum_{kk^{'}}\delta(k,k^{'})(T_{k^{'}}^{s})^{*}T_{k}^{s}\\
&=&\sum_{k,s}|T_{k}^{s}|^{2}\\
&=&\sum_{k,s}\frac{|\kappa|^{4}/|r|^{2}}{1+|r|^{2}-2|r|\cos(sE_{k}-\beta \Delta L)}|\left\langle k|m_{0}\right\rangle|\left\langle\psi_{k}^{s}|\phi_{in}^{s}\right\rangle|^2.
\end{array}
\end{equation}
By choosing an appropriate input state $\left|\phi_{in}^{s}\right\rangle$, $\sum_s|\left\langle k|m_{0}\right\rangle\langle\psi_{k}^{s}\left|\phi_{in}^{s}\right\rangle|^2$ could be independent on $k$. For instance, the input state in our experiment with the Gaussian mode $\left|m_{0}=0\right\rangle$ is prepared to the maximally mixed polarization state of $1/2(\left|\circlearrowleft\right\rangle\left\langle\circlearrowleft\right|+\left|\circlearrowright\right\rangle\left\langle\circlearrowright\right|)$, the total intensity $I_{\rm o}$ becomes
\begin{equation}
\begin{array}{rcl}
I_{\rm o}
&=&\sum_{k}\frac{|\kappa|^{4}/|r|^{2}}{1+|r|^{2}-2|r|\cos(sE_{k}-\beta \Delta L)}|\left\langle k|0\right\rangle|^2(|\left\langle\psi_{k}^{s}|\circlearrowleft\right\rangle|^2+|\left\langle\psi_{k}^{s}|\circlearrowright\right\rangle|^2)\\
&=&\sum_{k}\frac{|\kappa|^{4}/|r|^{2}}{1+|r|^{2}-2|r|\cos(sE_{k}-\beta \Delta L)}.
\end{array}\label{I0}
\end{equation}
On the other hand, the density of states related to volume V is defined as
\begin{equation}
    D(E)=\cfrac{1}{V}\sum_{k}\delta[E-E_{k}],
\end{equation}
where $E_{k}$ represents the energy band along momentum $k$.
In our experiment, only $sE_{k}=\beta \Delta L$ mainly contribute to the transmission intensity $I_{\rm o}(\beta \Delta L)$ in Eq. \ref{I0}. When $|r|\rightarrow 1$, $\frac{1}{1+|r|^2-2|r|x}$ will very close to the $\delta(x-1)$ function ($x=\cos(sE_{k}-\beta \Delta L)$). Thus, $I_{\rm o}$ can be approximated as 
\begin{equation}
\begin{array}{rcl}
I_{\rm o}(\beta \Delta L)&\approx&\Gamma\sum_{k}\delta(\beta\Delta L-sE_{k}),
\end{array}
\end{equation}
where $\Gamma$ is the normalised coefficient. Regarding $\Gamma$ as the volume V, $I_{\rm o}$ is denoted as density of the states 
\begin{equation}
I_{\rm o}(\beta \Delta L)=D(\beta \Delta L).
\end{equation}

\begin{figure*}[t!]
	\begin{center}
		\includegraphics[width=0.8 \columnwidth]{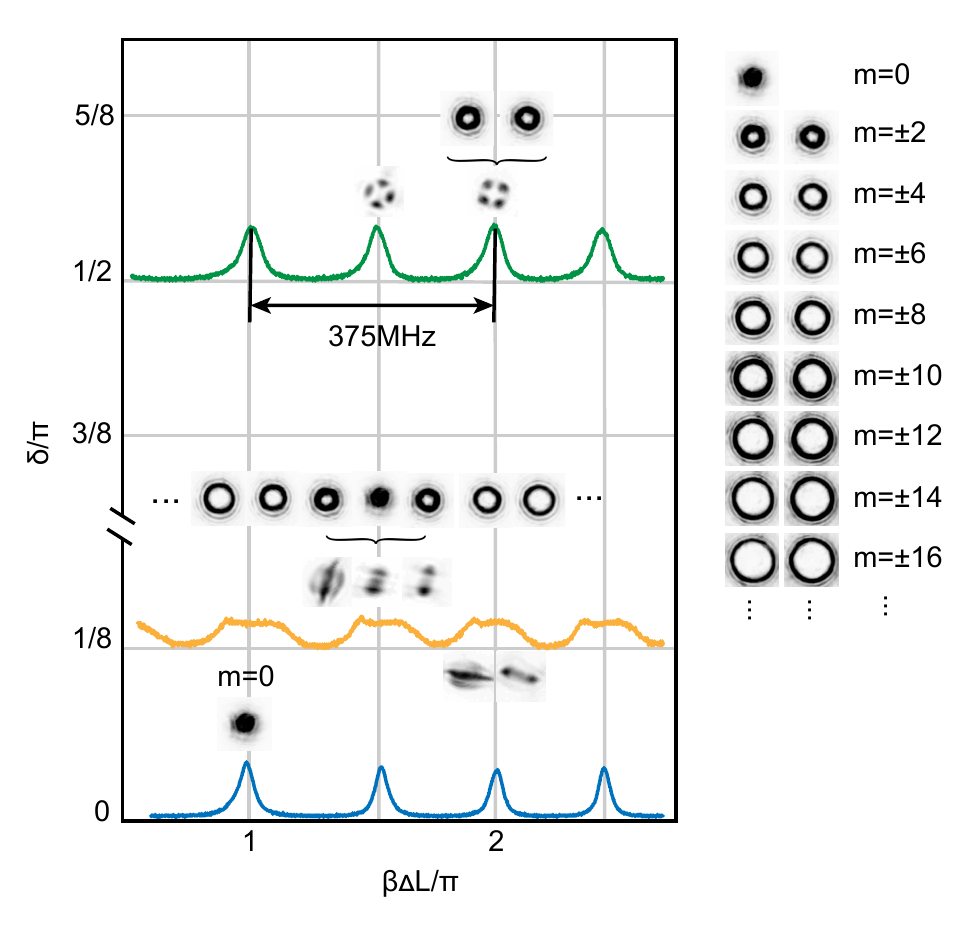}
		\caption{ \textbf{Experimental transmission modes at $\eta=\pi/4$.} The spectra are labled in blue ($\delta=0$), yellow ($\delta=1/8$) and green ($\delta=\pi/2$). The intensity distributions of transmitted photons, shown near the spectra, reveal the mixing of different angular momentum modes. The experimental spatial distributions of different angular momentum modes ($|m|=0\sim 16$) are shown on the right panel.}
	\label{modef}
	\end{center}
\end{figure*}

\section{The transmission modes of the cavity} 
Here we illustrate more details of the output modes of the cavity with $\eta=\pi/4$, which is shown in Fig. \ref{modef}. The input photons are on the Gaussian mode ($m=0$) with horizontal polarization. When $\delta=0$, the transverse mode of light is always kept in the Gaussian mode ($m=0$), but the polarization changes periodically. As a result, there are the splittings of the transmission peaks, which are twice that in the vacuum cavity. When $\delta>0$, the high order angular momentum modes begin mixed, and the spectra of the system getting more and more complicated, which satisfies the dispersion relation described in Eq. \ref{dispersion}. Especially when $\delta=\pi/2$, the transverse modes are restricted to the angular momentum modes with topological charge $|m|\leq 1$, which leads to the same spectrum of $\delta=0$. The images near the spectra represent the transmission intensity distributions, which are detected by a high-speed camera. These intensity distributions are formed by mixing the simple OAM ``ring" patterns. See the supplementary video for more details.

\section{Edge effect and disorder effect} 
\begin{figure*}[t!]
		\includegraphics[width=0.8 \columnwidth]{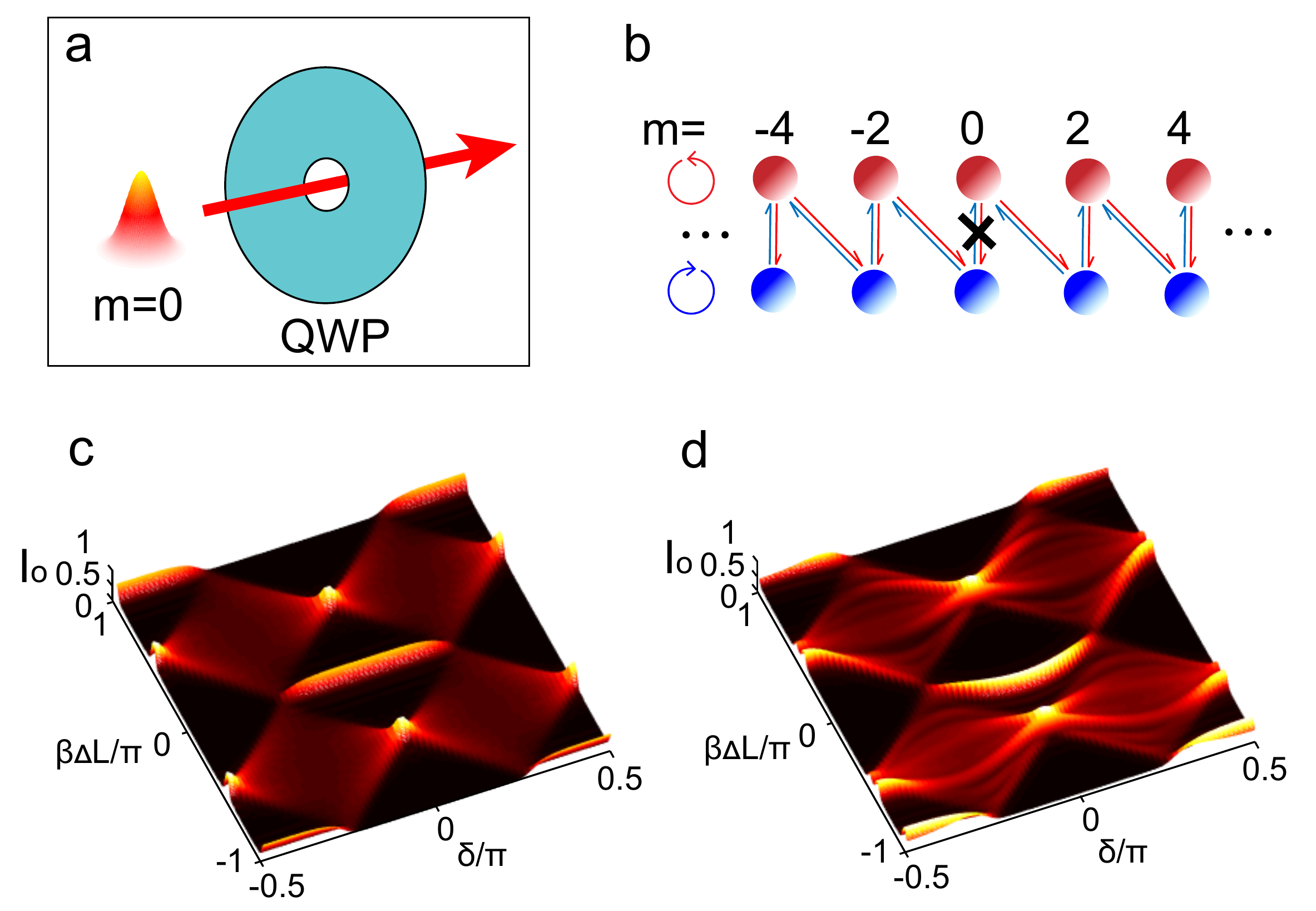}
		\caption{\textbf{The density of state (DOS) with edge effects.} \textbf{a}. A quarter wave plate (QWP) with a ping hole on the center is placed in the cavity. Only the Gaussian mode with $m=0$ can pass the hole and there is not coupling between different circular polarization. While for other modes passing the QWP, the coupling between different circular polarization occurs.
		\textbf{b}. The corresponding lattice with an edge at $m=0$. The lattice breaks into two parts.
		\textbf{c}.The numerical DOS without disorder. 
		\textbf{d}.The numerical DOS with disorder. 
		}
		\label{bound}
\end{figure*}

Edge states are topologically protected, an outstanding feature of topological physics.
Though the edge effect is weak in our current experiment, the edge states can be investigated in our platform by engineering the operation on different optical modes.
For instance, a QWP with a ping hole on its center, as shown in Fig. \ref{bound}a, can be used to realize such an operation. The radius of the ping hole is set to be 80 $\mu$m, which is the same as the waist radius of the Gaussian mode. Only the Gaussian mode with $m=0$ can pass the hole, and there is no coupling between different circular polarization. While for other modes passing the QWP, the coupling between different circular polarization occurs. The corresponding lattice has an edge at $m=0$, as shown in Fig. \ref{bound}b. The system breaks the symmetry at the lattice centre and it can then be viewed as a semi-infinite lattice, where the interface between the non-trivial topological bulk and ``vacuum" can support edge states. The numerical DOS without disorder is shown in Fig.\ref{bound}c, in which the edge states can be clearly seen at $E_{k}=0$ or $\pm\pi$.

The disorder is further introduced from the imperfect degeneracy of the cavity, which is given by a random phase $e^{i\Delta \theta_{m}}$ ($|\Delta \theta_{m}|<0.1\pi$) on each optical mode. Such kind of disorder corresponds to a distribution of energies around the main energy and makes the edge energy move to bulk bands. The simulated result is shown in Fig. \ref{bound}d. The edge state will merge into the bulk state with increasing the disorder strength.

\section{Direct measurement of the energy band spectrum} 
With the post-selected state $\left|k\right\rangle$, the transmission intensity becomes
\begin{equation}
I_{k}=\sum_{s}\sum_{k^{'}k^{''}}\langle k^{'}| k\rangle \left\langle k\right|\langle \psi_{k^{'}}^{s}|(T_{k^{'}}^{s})^{*}T_{k^{''}}^{s}|\psi_{k^{''}}\rangle |k^{''}\rangle=
\sum_{s}\sum_{k^{'}k^{''}}|T^s_{k}|^{2}\langle\psi_{k^{'}}^{s}|\psi_{k^{''}}^{s}\rangle\delta(k,k^{'})\delta(k,k^{''})
=\sum_{s}|T_{k}^{s}|^{2},
\end{equation}
which illustrates the distribution of $E_{k}$. By scanning the state $\left|k\right\rangle$, the energy band spectrum can be directly demonstrated.

\begin{figure*}[t!]
	\begin{center}
		\includegraphics[width=1 \columnwidth]{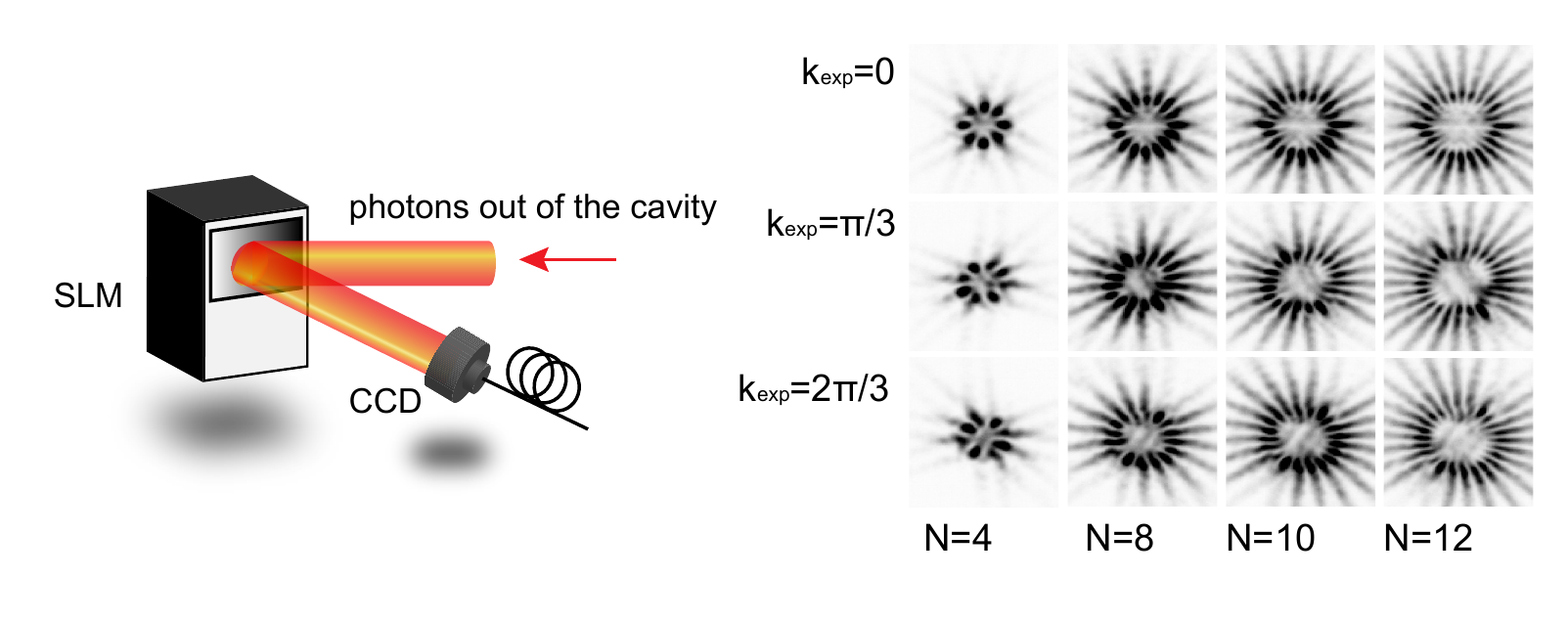}
		\caption{Experimental setup for post-selecting on the basis $\left|k_{\rm exp}\right\rangle\left\langle k_{\rm exp}\right|$. SLM: spatial light modulator; CCD: charge coupled device camera; (left).
		The photon distributions after post-selected by SLM with different settings of holograms (right).} 
			\label{modek}
	\end{center}
\end{figure*}

\section{Photon distributions after the modulation via SLM} 
The photon distributions after modulated by SLM of different $(k_{\rm exp},N)$ are shown in Fig. \ref{modek}. The number of the ``petals" of the interfered patterns is $2N$, while the patterns rotate with $k_{\rm exp}$.

\section{Direct measurement of the topological winding} 
Under the measurement basis $\left|k\right\rangle\left\langle k\right|\otimes(\sigma_{x},\sigma_{y},\sigma_{z})$, the output result gives,
\begin{equation}
(I^x_{k},I^y_{k},I^z_{k})=\sum_{k^{'}k^{''}}\langle k^{'}|k\rangle\left\langle k\right|\langle \psi_{k^{'}}^{s}|(T_{k^{'}}^{s})^{*}(\sigma_{x},\sigma_{y},\sigma_{z})T_{k^{''}}^{s}|\psi_{k^{''}}^{s}\rangle |k^{''}\rangle=\sum_{s}s(n_{x},n_{y},n_{z})|T_{k}^{s}|^{2},
\end{equation}
where 
\begin{equation}
\begin{array}{rcl}
\sigma_{x}&=&\left|H\right\rangle\left\langle H\right|-\left|V\right\rangle\left\langle V\right|,\\
\sigma_{y}&=&\left|A\right\rangle\left\langle A\right|-\left|D\right\rangle\left\langle D\right|,\\
\sigma_{z}&=&\left|\circlearrowright\right\rangle\left\langle \circlearrowright\right|-\left|\circlearrowleft\right\rangle\left\langle \circlearrowleft\right|,
\end{array}
\end{equation}
and 
\begin{equation}
\begin{array}{rcl}
\left|H\right\rangle&=&\dfrac{\left|\circlearrowright\right\rangle+\left|\circlearrowleft\right\rangle}{\sqrt{2}},\\
\left|V\right\rangle&=&\dfrac{\left|\circlearrowright\right\rangle-\left|\circlearrowleft\right\rangle}{\sqrt{2}},\\
\left|A\right\rangle&=&\dfrac{\left|\circlearrowright\right\rangle-i\left|\circlearrowleft\right\rangle}{\sqrt{2}},\\
\left|D\right\rangle&=&\dfrac{\left|\circlearrowright\right\rangle+i\left|\circlearrowleft\right\rangle}{\sqrt{2}}.
\end{array}
\end{equation}
The topological windings can be revealed by the variations of transmitted peaks.